\documentclass[11pt,a4paper]{article}
\usepackage{makeidx}
\usepackage{graphicx}
\usepackage{latexsym}
\usepackage{amsmath}
\usepackage{amsfonts}
\usepackage{verbatim}
\usepackage{nameref}
\usepackage{hyperref}
\usepackage{sectsty}
\usepackage{enumerate}

\usepackage{amssymb}

\usepackage{amsmath}
 \usepackage[dvips]{epsfig}
\usepackage{amsfonts}
\usepackage{amsmath,amscd}
\usepackage{amsbsy}
\usepackage{amscd}
\usepackage{float}
\usepackage{rotating}
\usepackage{xy} \xyoption{all} \xyoption{import} \xyoption{rotate}
\usepackage{pst-plot}
\usepackage{psfrag}
\usepackage{a4wide}
\newcommand{\bX}{{\bf x}}
\newcommand{\WW}{{W}}
\newcommand{\hatW}{\hat{W}}
\newcommand{\barW}{V}%\check{W}}

\newcommand{\alp}{\alpha}
\newcommand{\VV}{V}
\newcommand{\PP}{P}
\newcommand{\UU}{U}
\newcommand{\Oo}{\Omega}
\newcommand{\dO}{{\rm d} \Omega}
\newcommand{\dG}{{\rm d} \Gamma}
\newcommand{\dS}{{\rm d}S}
\newcommand{\dB}{{\rm d} \calB}

\newcommand{\Gu}{{\Gamma_{\chi}}}
\newcommand{\Gt}{{\Gamma_t}}
\newcommand{\St}{{S_t}}
\newcommand{\calB}{{\cal B}}
\newcommand{\uu}{{u}}

\newcommand{\be}{{\bf e}}

\newcommand{\xx}{{x}}
\newcommand{\la}{\label}
\newcommand{\half}{\frac{1}{2}}
\newcommand{\bt}{{\bf t}}

\newcommand{\bH}{{\bf H}}
\newcommand{\bE}{{\bf E}}
\newcommand{\bI}{{\bf I}}
\newcommand{\bC}{{\bf C}}
\newcommand{\bQ}{{\bf Q}}
\newcommand{\bF}{{\bf F}}
\newcommand{\bT}{{\bf T}}
\newcommand{\bn}{{\bf n}}
\newcommand{\bN}{{\bf N}}
\newcommand{\bx}{{\bf x}}

\newcommand{\sta}{{\rm sta}}
\newcommand{\vsig}{\zeta}% \varsigma}
\newcommand{\bxi}{\mbox{\boldmath$\xi$}}
 
\newcommand{\bchi}{\mbox{\boldmath$\chi$}}

\newcommand{\bsig}{\mbox{\boldmath$\sigma$}}
\newcommand{\bgamma}{\mbox{\boldmath$\gamma$}}
\newcommand{\btau}{{\mbox{\boldmath$\tau$}}}

\newcommand{\bbeta}{\mbox{\boldmath$\eta$}}

\newcommand{\calP}{{\cal P}}

\newcommand{\calE}{{\cal{E}}}
\newcommand{\calF}{{\cal{F}}}
\newcommand{\calS}{{\cal{S}}}
\newcommand{\barcalS}{\bar{\cal{S}}}
\newcommand{\calU}{{\cal{U}}}
\newcommand{\calT}{{\cal{T}}}
\newcommand{\calX}{{\cal{X}}}

\newcommand{\real}{\mathbb{R}}
\newcommand{\eb}{\begin{equation}}%[section]
 \newtheorem{thm}{Theorem}
 \newtheorem{Lemma}{Lemma}
 
  \newtheorem{definition}{Definition}
\newcommand{\ee}{\end{equation}}
\newcommand{\calW}{{\cal W}}

\newcommand{\Lam}{{\Lambda}}
  \newcommand{\bS}{{\bf S}}
    \newcommand{\barbS}{\bar{\bf S}}
  \newcommand{\barbchi}{\bar{\bchi}}

 \newtheorem{rem}{Remark}

\def \div{\mbox{div\hskip 1pt}}

\def \tr{\mbox{tr\hskip 1pt}}

 \vspace{-1.5cm}
\title{\bf {\Large Remarks  on Analytic Solutions   in Nonlinear  Elasticity and Anti-Plane Shear Problem}}
 \author{David Yang  Gao    \\[0.2cm]
 \small   Federation University Australia, Mt Helen, VIC 3353, Australia  \\
 \small Research School of Engineering, Australian National University, Canberra, Australia}
\date{}
\begin{document}
\maketitle

\begin{abstract}
 This paper revisits
 a well-studied anti-plane shear deformation problem formulated by Knowles in 1976
 and  analytical solutions
 in general nonlinear elasticity proposed by Gao since 1998.
 Based on minimum potential principle, a well-determined fully nonlinear system is obtained for isochoric deformation,
 which admits non-trivial states of finite anti-plane shear without ellipticity constraint.
 By using canonical duality theory,  a complete set of analytical solutions are  obtained for 3-D  finite deformation problems governed by
 generalized neo-Hookean model.
   Both global and local extremal solutions to the nonconvex variational problem  are identified by a triality theory.
 Connection between challenges in nonconvex analysis and NP-hard problems in computational science is revealed. 
 It is proved that the  ellipticity condition for general  fully  nonlinear boundary value problems  depends
  not only on differential operators, but also sensitively  on the external force field.
The homogenous hyper-elasticity  for   general anti-plane shear deformation  must be governed by the  generalized
neo-Hookean model.
 Knowles' over-determined system is simply due to a pseudo-Lagrange multiplier and two extra  equilibrium conditions
 in the plane. The constitutive condition  in his theorems is naturally satisfied with $b = \lambda/2$.
His ellipticity condition  is  neither   necessary nor sufficient
 for general  homogeneous  materials to admit nontrivial states of anti-plane shear.

\end{abstract}

 \section{Remarks on Nonconvex Variational Problem and Challenges}
Minimum total potential energy principle plays a fundamental role in continuum mechanics, especially for hyper-elasticity.
One important feature is that the equilibrium equations obtained (under certain regularity conditions)  by this principle are naturally compatible.
Therefore, instead of the local method adopted by Knowles \cite{knowles,knowles1}, the discussion of this paper begins from
the minimum potential variational problem  ($(\calP)$ for short):
  \eb
  (\calP): \;\; \min \left\{ \Pi(\bchi) = \int_{\calB} \WW(\nabla \bchi)  d\calB - \int_{S_t} \bchi \cdot \bt d S  | \;\; \bchi \in \calX_c \right\} , \label{eq-p}
  \ee
where the unknown deformation
  $\bchi (\bx) = \{ \chi_i (x_j) \} \in \calX_a$ is a vector-valued mapping
$\calB \subset \real^3 \rightarrow  \omega \subset  \real^3$
from a given material particle $\bx = \{ x_i\} \in \calB $ in the
undeformed body  to  a position vector  in the deformed configuration $ \omega$.
The body is fixed on the boundary $S_x \subset \partial \calB$, while on the remaining
boundary $S_t = S_x \cap \partial \calB$, the body is subjected to a given surface traction $\bt(\bx)$.
 In this paper, we assume
  \eb
  \calX_a = \{ \bchi \in \calW^{1,1} (\calB; \real^3) |  \;\; \det (\nabla \bchi) > 0 \;  \;\bchi(\bx) = 0 \;\; \forall \bx \in S_x  \},
  \ee
where   $\calW^{1,1} $ is the standard notation for Sobolev space, i.e. a  function space in which both
$\bchi$ and its weak derivative  $\nabla \bchi$ have a finite $L^1(\calB)$ norm.
Clearly, a function in $\calW^{1,1}$ is not necessarily  to be smooth, or even continuous.
For homogeneous hyperelastic body, the  strain energy $\WW(\bF)$   is assumed to be $C^1$ on its domain
  $\calF_a \subset  \real^{3 \times 3} $, in which certain necessary  {\em constitutive constraints} are included, such as
\eb\label{const}
  \WW(\bF) \ge 0 \;\; \forall \bF \in \calF_a, \;\; \WW(\bF) \rightarrow \infty  \mbox{ as } \| \bF \| \rightarrow \infty.
\ee
For incompressible materials, the condition $\det \bF = 1$ should be included.
Finally,  $\calX_c = \{ \bchi \in \calX_a | \; \nabla \bchi \in \calF_a \}$ is the {\em kinetically admissible space},
which is nonconvex due to  nonlinear   constraints such as  $\det (\nabla \bchi)  > 0 $.
 Also, the stored energy $\WW(\bF)$ is  in general nonconvex.
   Therefore,   the nonconvex variational problem $(\calP)$ has usually multiple local optimal solutions.

Let $\calX_b \subset \calX_c$ be a subspace with two additional conditions
\eb\label{cw}
\calX_b = \{ \bchi \in \calX_c | \; \; \bchi \in C^2(\calB; \real^3), \;\;  \WW(\bF(\bchi) )  \in C^2(\calF_a; \real) \},
\ee
the criticality condition $\delta \Pi(\bchi; \delta \bchi) = 0 \;\; \forall \delta \bchi \in \calX_b$ leads to  a nonlinear boundary-value problem
\eb
(BVP): \;\; \; \left\{
\begin{array}{l}
 - \nabla \cdot \bsig (\nabla \bchi) = 0 \;\; \mbox{ in } \calB, \\
\bN \cdot \bsig (\nabla \bchi) = {\bf t} \;\; \mbox{ on } S_t  ,\;\;   \bchi  = 0 \;\; \mbox{ on } S_x
 \end{array}\right.
 \label{eq-ebp}
 \ee
 where,  $\bN \in \real^3$ is a unit vector normal to $\partial \calB$,
  and $\bsig(\bF)$  is the first Piola-Kirchhoff   stress (force per unit undeformed
area),  defined by
\eb
\bsig   = \nabla \WW(\bF) , \;\; \mbox{ or } \;  \sigma_{ij}  = \frac{\partial \WW(\bF)}{\partial F_{ij} }, \;\;
i,j = 1,2,3
\ee
which is also a two-point tensor.

\begin{rem}[KKT Conditions,  Isochoric Deformation,  pseudo-Lagrange Multiplier] %$\;$\newline
{\em Strictly speaking, there is an inequality constraint in $\calX_c$, i.e. the admissible deformation condition
 $\det (\nabla\bchi)  > 0$.
 According to the mathematical theory of non-monotone variational inequality,% \cite{liu-gao},
  in addition to the equilibrium equations in
  $(BVP)$, we have the following KKT conditions
 \eb\label{kkti}
p  \le  0, \;\; \det (\nabla \bchi) > 0, \;\; p \det (\nabla \bchi) = 0
\ee
where  $p$ is a Lagrange multiplier and $p \le 0$ is called the condition of constraint qualification.
The equality  $p \det (\nabla \bchi) = 0 $ is the well-known {\em complementarity condition}
 in  variational inequality theory, by  which we must have $p = 0$
 in order   to guarantee  the   inequality constraint
 $\det (\nabla \bchi) > 0$. Therefore, this constraint  is actually not active to the   problem $(\calP)$.
 Such an inactive constraint is not  a variational constraint.

For incompressible deformation,  the inequality condition  $\det (\nabla \bchi) > 0 $ in $\calX_c$ should be
   replaced by an equality constraint $\det (\nabla \bchi) = 1$. In this case, $(\calP)$ is a constrained variational problem.
   The   KKT conditions (\ref{kkti}) should be replaced  by (see \cite{lat-gao-ol15})
 \eb
  p \neq 0, \;\; \det \bF(\bchi) = 1, \;\; p  (\det \bF(\bchi)  - 1) = 0 \label{eq-compf}.
  \ee
  and we must have   $p(\bx) \neq 0 \mbox{ for a.e. } \bx \in \calB$
  in order to ensure $\det \bF(\bchi)  - 1 = 0$.
  The associated  $(BVP)$ should be
  \eb \label{eq-geqi}
(BVP)_p: \;\; \left\{
\begin{array}{l}
  - \nabla \cdot \bsig(\nabla \bchi, p)  =   0, \;\; \det (\nabla \bchi) = 1 \;\; \mbox{ in } \calB , \\
  \bN \cdot  \bsig(\nabla \bchi, p)  = \bt \;\; \mbox{ on } S_t, \;\; \bchi  = 0 \;\; \mbox{ on } S_x.
\end{array}
\right.
\ee
in which,
 $    \bsig (\bF, p) = \nabla \WW(\bF) - p  \bF^{-T} $,
  where   $\bF^{-T} = (\bF^T)^{-1}$.
In this case, we have  two variables $(\bchi, p)$  and two equations in $\calB$,  thus, the problem $(BVP)_p$ is a well-defined system.

For  isochoric (i.e.  volume preserving)  deformation, say the  anti-plane shear problems,
the condition  $\det \bF = 1$ is trivially satisfied and the  complementarity condition
$p(\det \bF - 1) \equiv 0  \;\; \forall p(\bx) \neq 0 \;\; a.e. $ in $\calB$. In this case, the trivial condition $\det \bF = 1$ is not
a variational constraint for $(\calP)$ and  the arbitrary parameter   $p(\bx)$ is not  an  unknown  variable for $(BVP)_p$.
Otherwise, the $(BVP)_p$ is an over-determined system. This fact in KKT theory  is important for understanding Knowles'   anti-plane shear problem.
Such a parameter for trivial condition can be called {\em pseudo-Lagrange multiplier}.

  Physically speaking, the hydrostatic pressure  $p$ is not necessary to be  zero
  even for  isochoric deformations.
There are many  examples  in the literature,
  see  the celebrated book by  Ogden \cite{ogden} as well as
  many famous papers by Rivlin on volume-preserving deformations of isotropic materials (simple shear, torsion, flexure, etc.)\footnote{Personal communications with David Steigmann, Ray Ogden, and C. Horgan}.\hfill $\blacksquare$
}
\end{rem}

\begin{rem}[Convexity, Multi-Solutions, and NP-Hard Problems]$\;$\newline
{\em
The stored energy $\WW(\bF)$ in nonlinear elasticity is generally nonconvex. It turns out that the fully nonlinear $(BVP)$
could have  multiple solutions $\{\bchi_k(\bx) \}$ at each material point $\bx\in \calB_s \subset \calB$.
As long as the continuous domain  $\calB_s \neq \emptyset$, this solution set $\{ \bchi_k(\bx) \}$ can form  infinitely many solutions
to $(BVP)$ even $\calB \subset \real$. It is impossible to use traditional convexity and ellipticity conditions to identify global minimizer
among all these local solutions. 
 Gao and Ogden discovered in \cite{gao-ogden-qjmam} that
for certain given external force field, both global and local extremum solutions are nonsmooth and can't be obtained
by  Newton-type numerical methods. Therefore, Problem $(\calP)$ is much more difficult than $(BVP)$.
 In computational mechanics, any direct numerical method  for solving   $(\calP)$ will lead  to
a nonconvex minimization problem.
Due to the lack of global optimality condition, it is a well-known challenging task to solve nonconvex minimization problems by traditional methods. Therefore,  in  computational sciences most nonconvex minimization problems are  considered to be NP-hard (Non-deterministic Polynomial-time hard)  \cite{gao-bridge}.

Direct methods for solving nonconvex  variational problems   in finite elasticity have been
studies extensively during the last fifty years and  many generalized
convexities, such as poly-, quasi- and rank-one convexities, have been proposed.
For a  given function  $W:\calF_a \rightarrow \real$, the
following statements  are well-known (see  \cite{sch-neff})\footnote{It was proved recently that rank-one convexity  also implies polyconvexity for
isotropic, objective and isochoric elastic
energies in the two-dimensional case \cite{neff15}.}:
\[
 \mbox{    convex $\Rightarrow   \mbox{ poly-convex }  \Rightarrow
  \mbox{ quasi-convex }   \Rightarrow   \mbox{ rank-one convex}$.}
  \]

Although the generalized convexities have been well-studied  for general function $\WW(\bF)$ on matrix space
$\real^{m\times n}$, these mathematical concepts  provide only necessary conditions for local minimal solutions, and 
can't be   applied  to general finite deformation problems.
In reality, the stored energy $\WW(\bF)$ must be nonconvex in order to model real-world phenomena, such as
post-buckling and  phase transitions etc.
Strictly speaking,  due to certain necessary  constitutive constraints such as  $\det \bF > 0$  and objectivity etc,
even the domain $\calF_a$ is not convex, therefore, it is not appropriate to discuss convexity of the stored energy $\WW(\bF)$ in general nonlinear elasticity.
  How to identify global optimal solution has been a fundamental challenging problem in nonconvex analysis and computational science.
\hfill $\blacksquare$ }
\end{rem}

\begin{rem}[Canonical Duality,  Gap Function, and Global Extremality]$\;$\\
{\em
The objectivity is a necessary constraint   for any hyper-elastic model.
 A real-valued function  $\WW:\calF_a \rightarrow \real$  is  objective  iff
%\[ W(\bF) = W(\bQ \bF) \;\; \forall \bF \in \calF_a, \;\; \forall \bQ \in \mathbb{O}_+^3 = \{ \bQ \in \real^{3\times 3} |\;\; \bQ^T = \bQ^{-1}, \;\; \det \bQ = 1\} \]  or equivalently, iff
 there exists a function $\UU(\bC)$ such that
$
W(\bF)  = \UU(\bF^T \bF) \;\; \forall \bF \in \calF_a.
 $
By the fact that the right Cauchy-Green tensor $\bC$ is an objective measure on a convex domain
$\calE_a = \{ \bC \in \real^{3\times 3} | \;\; \bC = \bC^T, \;\; \bC \succ 0 \}$, it is possible and natural
to discuss the convexity of
$\UU(\bC)$.
This fact lays a foundation for the canonical duality theory \cite{gao-dual00}, which was developed from
  Gao and Strang's original work  in 1989 \cite{gao-strang89a} for general nonconvex/nonsmooth  variational problems in finite deformation theory.
  The key idea of this theory is assuming the existence of a  geometrically admissible (objective) measure
  $\bxi=\Lambda(\bF)$ and a canonical function $\VV(\bxi)$ such that the following {\em  canonical transformation}  holds
 \eb
 \bxi = \Lambda(\bF):\calF_a \rightarrow \calE_a   \;\;  %\VV(\bxi):\calE_a \rightarrow \real \mbox{ is convex }  \;
 \Rightarrow \; \; \WW(\bF) = \VV(\Lambda(\bF)). \label{eq-lam}
 \ee
 A real-valued function $\VV:\calE_a \rightarrow \real$ is called  canonical if the duality relation
 $\bxi^* = \nabla \VV(\bxi) : \calE_a \rightarrow \calE^*_a$ is one-to-one and onto \cite{gao-dual00}.
 This canonical duality is necessary for modeling natural phenomena. 
  Gao and Strang discovered that  the  directional  derivative $\Lam_t (\bF)= \delta\Lam(\bF) $  is adjoined with
the  equilibrium operator,  while its complementary operator $\Lam_c(\bF) = \Lambda(\bF) - \Lam_t(\bF) \bF $   leads to
 a so-called {\em    complementary gap function}, which recovers  duality gaps in traditional analysis and  provides a sufficient  condition for identifying both global and local extremal solutions
 \cite{gao-dual00,gao-bridge}.\hfill $\blacksquare$
 }\end{rem}

The canonical duality  theory has been applied for solving a large class of nonconvex, nonsmooth, discrete problems
  in multidisciplinary fields of nonlinear  analysis,   nonconvex  mechanics, global optimization, and computational sciences, etc.
  A comprehensive review is given recently in \cite{gao-bridge}.
 The main goal of this paper is to show author's  recent  analytical  solutions \cite{gao-cmt15}
 are valid for general anti-plane shear problems and can be easily generalized for solving finite deformation problems governed by generalized neo-Hookean materials. While the constitutive constraints in Knowles' over-determined system \cite{knowles}
  are not necessary, the hydrostatic pressure $p= p(x_1,x_2)$  is independent of $x_3$ and can't be considered as
  a variational variable.
  Some insightful  results  are obtained on ellipticity condition in  nonlinear analysis.

\section{Complete Solutions to Generalized Neo-Hookean Material}
Since the right Cauchy-Green strain $\bC = \bF^T \bF$ is  an objective tensor,  its three
  principal invariants
\begin{equation}
I_1(\bC)=\tr\mathbf{C},\quad  I_2 (\bC)= \half [ (\tr \bC)^2 - \tr(\bC)^2 ]
 ,\quad I_3 (\bC)=\det\mathbf{C}
\end{equation}
are also objective functions of $\bF$. Clearly, for  isochoric   deformations we have $ I_3 (\bC) = 1 $.
The elastic body is said to be  {\em  generalized neo-Hookean material} if  the stored energy depends only on  $I_1$,
  i.e. there exists a function $\barW(I_1) $
 such that $W(\bF) =  \barW(I_1(\bC(\bF)) ) $.
Since $I_1 = \tr(\bF^T\bF) > 0 \;\; \forall \bF \in \calF_a$, the   domain of $\barW(I_1)$
   is a   convex (positive)  cone $\calE_a =   \{ x \in \real| \; x > 0 \}$,
   it is possible to discuss  the  convexity of $\barW(I_1)$.
  Furthermore, we assume that $\barW(I_1)$ is a $C^2(\calE_a )$ canonical function.
   Then the canonical transformation (\ref{eq-lam}) for the generalized neo-Hookean model is
\eb
 \xi = \Lam(\bF) = \tr(\bF^T \bF) : \calF_a \rightarrow \calE_a, \;\;
 \WW(\bF) = \barW(\xi(\bF)) .% \;\; \nabla \VV:\calE_a \rightarrow \calE^*_a  \mbox{ is one-to-one and onto} .
\ee
For a  given external force $\bt(\bx) $ on $\St$, we introduce
  a {\em statically admissible space}
\eb
{\calT}_a = \{ { \bT } \in {\cal C}^1[\calB ; \real^{3\times 3}] \; | \;\; \nabla \cdot \bT = 0 \;\; \mbox{ in } \calB, \;\;
 \bN \cdot \bT= \bt \mbox{ on } S_t \}.
 \ee
Thus for any given $\bT \in \calT_a$, %the total potential $\Pi(\bchi)$ can be equivalently reformed as
%\eb  \Pi(\bchi) = \int_{\calB} \PP(\nabla \bchi)\dB, \;\; \PP(\bF) =  \WW(\bF) - \bF : \bT\ee
the primal problem $(\calP)$ for the generalized neo-Hookean material can be written in following canonical form
 \eb
 (\calP)_g : \;\; \min \left\{ \Pi(\bchi) = \int_{\calB} \PP(\nabla \bchi) \dB , \;\;   | \;\;
 \forall \bchi \in \calX_c \right\},
 \ee
where $\calX_c = \{ \bchi \in \calX_a| \;  \Lambda(\nabla \bchi) \in \calE_a \}$ and the integrand
$\PP:\calF_a \rightarrow \real$ is defined by
\eb
 \PP(\bF) = \barW(\Lambda(\bF)) - \tr(\bF \bT ).
 \ee

The criticality condition for this canonical variational problem leads to the following canonical
boundary value problem
\eb
(BVP)_g: \;\;
\nabla \cdot [2 \vsig (\nabla \bchi)] = 0 \;\; \mbox{ in } \calB, \;\; \bN \cdot [2 \vsig (\nabla \bchi)]  = \bt \; \mbox{ on } \St, \;\; \bchi = 0 \;\; \mbox{ on } S_x
\ee
which are identical to $(BVP)$ since   $\nabla \WW(\bF) = 2 \bF   [\nabla \barW(\xi)] = 2 \vsig  \bF $.
To solve  this fully nonlinear boundary value problem is very difficult for direct methods, but easy for the canonical duality theory.

By the canonical assumption of $\barW(\xi)$, the duality relation   $\vsig = \nabla \barW(\xi) : \calE_a \rightarrow \calE^*_a$ is invertible.
The complementary energy  can be defined uniquely  by the
Legendre transformation
\eb
\barW^*(\vsig) = \{\xi \vsig - \barW(\xi) | \; \vsig = \nabla \barW(\xi) \}.
\ee
 Clearly, the function   $\barW:\calE_a \rightarrow \real$ is   canonical if and only if the following canonical duality relations hold on $\calE_a \times \calE^*_a$
\eb
\vsig = \nabla \barW(\xi) \;\; \Leftrightarrow \;\; \xi = \nabla \barW^*(\vsig)  \;\; \Leftrightarrow
\;\;
\barW(\xi) + \barW^*(\vsig) = \xi \vsig.
\ee
Using $\barW(\xi) = \xi\vsig - \barW^*(\vsig)$, the nonconvex function
$\PP(\bF)$ can be written as the so-called total complementary function on $\calX_a \times \calE^*_a $
\eb
\Xi(\bF, \vsig) =  \Lam(\bF) \vsig - \barW^*(\vsig) - \tr(\bT \bF) .
\ee
The canonical dual function  can be obtained by the canonical dual transformation:
\eb
\PP^d(\vsig) = \{ \Xi(\bF, \vsig) | \;\; \nabla_{\bF} \Xi(\bF, \vsig) = 0 \}
=     - \barW^*(\vsig) - \frac{1}{4} \vsig^{-1} \tau^2 , \;\;\;     \tau^2 =   \tr(\bT^T  \bT ) .
  \ee
 Thus,  the pure complementary energy  principle, first proposed   in 1998 \cite{gao-ima98},
 leads to the following canonical dual variational problem
  \eb
 (\calP^d)_g: \;\;\;\; \sta \left\{  \Pi^d(\vsig) =   \int_{\calB} \PP^d(\vsig) \dB \; | \;\; \vsig \in \calS_a \right\} ,
  \ee
 where    $\sta \{ \Pi^d(\vsig)| \;\vsig \in \calS_a\}$ stands for finding stationary point of $\Pi^d(\vsig)$
 on  the canonical dual feasible space  $ \calS_a = \{ \vsig \in \calE^*_a | \; \vsig^{-1} \tau^2 \in L^1(\calB) \}$.

 Since the canonical dual variable $\vsig$ is a scalar-valued function, the criticality condition $\delta \Pi^d (\vsig) = 0 $ leads to  a
so-called   {\em canonical dual algebraic equation} (see \cite{gao-dual00}):
  \eb\label{cda}
  4 \vsig^2  \nabla \barW^*(\vsig)  =  \tau^2
  \ee
Note that  $\nabla \barW^*(\vsig):\calE^*_a \rightarrow \calE_a$ is also one-to-one and onto,
 this equation has at least one solution for any given $\tau^2 = \tr(\bT^T\bT) \ge 0 $ and $\vsig = 0 $ only if $\tau = 0$. Therefore, $\PP^d(\zeta)$ is well-defined.
Due to the nonlinearity, the solution may not be unique \cite{gao-cmt15}.
By the pure complementary energy principle proposed by  Gao in 1999
(see \cite{gao-dual00}), we have
\begin{thm}[Pure Complementary Energy Principle]
For any given nontrivial $\bt \neq 0$ such that  $\bT \in \calT_a$,  (\ref{cda}) has at least one solution    $\vsig_k \neq 0$,
 the deformation vector defined by
   \eb\label{eq-solu}
   \bchi_k (\bx) = \half \int_{\bx_0}^\bx  \vsig_k^{-1}  \bT\cdot \mbox{d}\bx
   \ee
   along any path from $\bx_0 \in  S_x$ to $\bx \in \calB$ is a critical point of $\Pi(\bchi)$ and
   $\Pi(\bchi_k) = \Pi^d( \vsig_k) $.
   \end{thm}

%   Dually, if $\bchi_k$ is a critical point of $\Pi(\bchi)$, then it must be the form of (\ref{eq-solu})   such that
%    $\vsig_k = \frac{1}{6} \tr[\bT  \bF^{-1}(\bchi_k)] $ is a critical point of $\Pi^d(\vsig)$ for   a  given $\bT \in \calT_a$. }\\

 This principle shows that by the canonical dual transformation, the nonlinear partial differential equation
 in $(BVP)$ for generalized neo-Hookean model can be converted to an algebraic equation (\ref{cda}),
 which can be solved   to obtain a complete set of solutions (see \cite{gao-cmt15,gao-haj}).
% By the fact that this principle solved a well-known debate in nonlinear elasticity, it is known as the Gao principle  \cite{li-gup}.

  Since $\calS_a $ is nonconvex, in order to identify global and local optimal solutions, we need
 the following convex subsets
\eb
\calS_a^+ = \{ \vsig \in \calS_a| \; \zeta   > 0 \}, \;  \;\;
\barcalS_a^+ = \{ \vsig \in \calS_a| \; \zeta  \ge 0
 \}, \;\; \calS_a^- = \{ \vsig \in \calS_a| \;  \zeta  < 0
 \}.
\ee
Then by  the canonical duality-triality theory developed in \cite{gao-dual00} we have   the following theorem.

\begin{thm}\label{thm-1}
Suppose that $\VV:\calE_a \rightarrow \real$ is convex and  for a  given $\bT \in \calT_a$ such that $\{\vsig_k\}$ is a solution set to (\ref{cda}),
 $\bF_k =  \half  \vsig^{-1}_k \bT  $, and
 $\bchi_k$ is defined by (\ref{eq-solu}),
 we have the following statements.
\begin{verse}
1. If $\vsig_k \in \barcalS_a^+$,
then  $\nabla^2 \WW(\bF_k) \succeq 0$ and   $\bchi_k$ is a global minimal solution to $(\calP)_g$.\\

2. If $\vsig_k \in \calS_a^-$ and $\nabla^2  \WW(\bF_k)\succ 0$, then $\bchi_k$ is a local minimal solution to  $(\calP)_g$.\\

3. If $\vsig_k\in \calS_a^-$ and $\nabla^2  \WW(\bF_k) \prec 0$, then $\bchi_k$ is a local maximal solution to  $(\calP)_g$.
\end{verse}

If $\{\vsig_k\} \subset  \barcalS^+_a $,  then $\{\bchi_k\}$ is a convex set.
The problem $(\calP)_g$  has a unique solution if $\{\vsig_k\} \subset \calS^+_a$. % $\barbchi_k \in \calX_s$.
\end{thm}
{\bf Proof}.
 By using  chain rule for $\WW(\bF ) = \barW(\xi(\bF))$ we have $\nabla \WW(\bF) = 2 \bF   [\nabla \barW(\xi)] = 2 \vsig  \bF $, and
 \eb \label{eq-hatw}
   \nabla^2 \WW(\bF ) = 2 \vsig  \bI\otimes \bI   + 4  h(\xi)  \bF   \otimes \bF  ,
 \ee
% or in  the form of components \eb
% \frac{\partial^2 W(\bF)}{\partial F^i_\alp \partial F^j_\beta} =2  \delta^{ij}  S_{\alp\beta} + 4
% \sum_{\theta, \nu = 1}^3  F^i_\theta  H_{\theta \alp\beta \nu} F^j_\nu, \ee
  where $\mathbf{I}$ is an identity tensor in $\real^{3\times 3}$,  $h(\xi) =   \nabla^2 \barW(\xi) \ge 0 $ due to the convexity of   $\barW$
 on $ {\calE_a}$.
 Therefore, $ \nabla^2 \WW(\bF_k )  \succeq 0$ if $\vsig_k \in \barcalS^+_a$.

 To prove $\bchi_k$ is a global minimizer of $(\calP)$, we  follow  Gao and Strang's work in 1989 \cite{gao-strang89a}. By the
  convexity of $\barW(\xi)$ on its convex domain $\calE_a$, we have
\eb\label{eq-xic}
\barW(\xi) - \barW(\xi_k) \ge   (\xi - \xi_k)  \vsig_k \;\;  \;\; \forall \xi,\; \xi_k \in \calE_a,\;\; \vsig_k = \nabla \barW(\xi_k).
\ee
For any given variation $\delta \bchi$, we let $\bchi= \bchi_k + \delta \bchi$.
Then we have \cite{gao-strang89a}
\eb\label{lamtc}
\Lambda(\nabla\bchi) = \tr[(\nabla \bchi)^T    (\nabla \bchi)]  = \Lambda(\nabla\bchi_k) + \Lam_t (\nabla \bchi_k)   (\nabla \delta  \bchi)    - \Lambda_c(\nabla \delta \bchi),
\ee
where $\Lam_t(\bF) \delta \bF = 2 \tr[ \bF^T  (\delta \bF) ]$ and  $
\Lambda_c(\delta \bchi)  =  - \Lambda ( \delta  \bchi)$. Clearly, $\Lam(\bF) = \Lam_t(\bF) \bF + \Lam_c(\bF)$.
Then combining  the inequality (\ref{eq-xic}) and (\ref{lamtc}), we have
\begin{eqnarray}
\Pi(\bchi) - \Pi(\bchi_k) &\ge& \int_\calB 2 \vsig_k \tr[ (\nabla \bchi_k)^T   (\nabla \delta \bchi)]   \dB
 - \int_\St \delta \bchi \cdot \bt \dS  +  \int_\calB  \vsig_k \tr [(\nabla \bchi)^T  (\nabla \bchi) ] \dB\nonumber \\
 &=& \int_\calB    [2\vsig_k  (\nabla \bchi_k)  - \bT ] : (\nabla \delta \bchi)   \dB
 + G_{ap}(\delta\bchi, \vsig_k)  \;\; \forall  \bchi, \; \delta \bchi \in \calX_c
\end{eqnarray}
for any given $ \bT \in \calT_a$, where
\eb
G_{ap}(\bchi, \vsig) = \int_\calB -\Lambda_c(\nabla\bchi) \vsig \dB = \int_\calB  \vsig \tr [(\nabla \bchi)^T  (\nabla \bchi) ] \dB
\ee
is the {\em complementary gap function} introduced in \cite{gao-strang89a}.
If $\bchi_k$ is a critical point of $\Pi(\bchi)$, then we have
\[
\int_\calB   [2 (\nabla \bchi_k) \vsig_k - \bT ] : (\nabla \delta \bchi)   \dB = 0 \;\; \forall  \delta \bchi \in \calX_c, \;\;
\forall \bT \in \calT_a
\]
%$\nabla \bchi_k = \half \bT \cdot \bS_k^{-1}$.
%and
%$\nabla \UU^*(\bS_k) = \frac{1}{4} [\bT \bS_k^{-1}]^T \cdot [\bT \bS_k^{-1}] = (\nabla \bchi_k)^T \cdot (\nabla \bchi_k)$.
Thus, we have
$
\Pi(\bchi) - \Pi(\bchi_k) \ge G_{ap}(\delta \bchi, \vsig_k) \ge 0 \;\; \forall  \delta \bchi \in \calX_c  \;  \mbox{ if } \vsig_k \in \barcalS_a^+.
$
This shows that $\bchi_k$ is a global minimizer of $(\calP)_g$.

To prove the local extremality, we replace $\bF_k$ in  (\ref {eq-hatw}) by  $\bF_k = \half \vsig^{-1}_k \bT$ such that
 \eb
 {\bf G} (\zeta_k) =   \nabla^2 \WW(\bF_k) = 2 \vsig_k \bI\otimes \bI   + \vsig_k^{-2}  h(\xi_k)  \bT   \otimes \bT  ,
 \ee
where $\xi_k = \nabla \barW^*(\vsig_k)$. Clearly, for a given $\bT \in \calT_a$ such that $\vsig_k \in \calS^-_a$,
 the Hessian $ \nabla^2 \WW(\bF_k)$ could be either positive or negative definite.
The total potential $\Pi(\bchi_k)$ is locally convex if the {\em Legendre condition}
$ \nabla^2 \WW(\nabla \bchi_k) \succeq 0$ holds, locally concave
 if   $ \nabla^2 \WW(\nabla \bchi_k) \prec 0$.
 Since $\bchi_k$ is  a global minimizer when $\vsig_k \in \barcalS^+_a$,
 therefore, for   $\vsig_k \in \calS^-_a$, the stationary solution $\bchi_k$ is  a local  minimizer  if
 $ \nabla^2 \WW(\nabla \bchi_k) \succeq 0$ and, by the triality theory\cite{gao-dual00,gao-bridge},
 $\bchi_k$ is a biggest local maximizer if
 $ \nabla^2 \WW(\nabla \bchi_k) \prec 0$.

 If $\{\vsig_k \} \subset  \barcalS^+_a $, then  all    the solutions $\{ \bchi_k \} $ are global minimizers and
 form a convex set. Since $\PP^d(\zeta)$ is strictly concave on the open convex set $\calS^+_a $,
the condition  $\{\vsig_k \} \subset \calS^+_a$ implies the unique solution of (\ref{cda}).
 In this case, both problems $(\calP)$ and  $(BVP)$  have at most one  solution.
 \hfill $\Box$
\begin{thm}[Triality Theory]
For any given $\bT \in \calT_a$,  let  $\zeta_k$ be    a critical point of   $ (\calP^d)_g$,
the vector   $\bchi_k$ be  defined by (\ref{eq-solu}), and
 $ \calX_o \times \calS_o  \subset
 \calX_c \times \calS^-_o$   a neighborhood\footnote{The neighborhood $\calX_o$ of $\bchi_k$ in the canonical duality theory  means that $\bchi_k$ is the only one critical point of $\Pi(\bchi)$ on $\calX_o$ (see \cite{gao-dual00}).}
 of $ (\bchi_k, \zeta_k)$.

If $\zeta_k \in \calS^+_a$, then
 \eb
 \Pi(\bchi_k) = \min_{\bchi \in \calX_c} \Pi(\bxi) = \max_{\vsig \in \calS^+_a} \Pi^d (\vsig) = \Pi^d(\vsig_k).
 \ee

If $\zeta_k \in \calS^-_a$ and  $ {\bf G} (\zeta_k) \succ 0$, then
\eb
 \Pi(\bchi_k) = \min_{\bchi \in \calX_o} \Pi(\bxi) = \min_{\vsig \in \calS_o} \Pi^d (\vsig) = \Pi^d(\vsig_k).
 \ee

 If $\zeta_k \in \calS^-_a$  and $ {\bf G} (\zeta_k) \prec  0$,
 then
 \eb
 \Pi(\bchi_k) = \max_{\bchi \in \calX_o} \Pi(\bxi) = \max_{\vsig \in \calS_o} \Pi^d (\vsig) = \Pi^d(\vsig_k).
 \ee
 \end{thm}

This theorem   shows that the triality theory can be used to identify
both global and local extremum solutions to  the  variational problem $(\calP)_g$ and
the nonconvex minimum variational problem $(\calP)_g$ is canonically equivalent to
the following concave maximization problem over an open  convex
set $\calS^+_a$, i.e.
 \eb
 (\calP^\sharp)_g: \;\;\;\; \max \left\{  \Pi^d(\vsig) =   \int_{\calB} \PP^d(\vsig) \dB \; | \;\; \vsig \in \calS^+_a \right\} ,
  \ee
  which is much easier to solve for obtaining  global optimal solution of $(\calP)_g$.

  \section{Generalized Quasiconvexity, G-Ellipticity,  and Uniqueness}

The  ellipticity is a classical concept originally from linear partial differential systems,
 where  the deformation is a scalar-valued function $\chi:\calB \rightarrow \real$ and  stored energy is a  quadratic function
$\WW(\bgamma) = \half \bgamma^T  {\bH}  \bgamma $  of $\bgamma = \nabla \chi \in \real^3$.
The linear operator
\[
L[\chi] = - \nabla \cdot  [\bH   (\nabla \chi)]  =  - [h_{ij} \chi_{,j}]_{,i}
\]
 is called elliptic if
 $\bH = \{ h_{ij} \} $ is positive definite.  In this case, the function  $\PP(\bgamma ) = \WW(\bgamma)  - \bgamma^T \btau$
 is convex  and its level set  is an ellipse for any given $\btau \in \real^3$.
This concept has been extended  to   nonlinear analysis.
The fully nonlinear  partial differential  equation
 in $(BVP)$ (\ref{eq-ebp}) is called elliptic   if
the stored energy $\WW(\bF)$ is rank-one convex.
In the case   $W \in {C}^2$, the rank-one convexity is equivalent to the Legendre-Hadamard (LH) condition:
  \eb \label{eq-LH}
({\bf a} \otimes {\bf a} ) :  \nabla^2\WW(\bF) : ({\bbeta}  \otimes {\bbeta} ) \ge 0 \;\; \forall {\bf a}, \bbeta \in \real^3,
 \;\; \forall \bF \in \calF_a \subset \real^{m\times n}.
\ee
The $(BVP)$ is called strong elliptic if the inequality holds strictly. In this case, $(BVP)$ has at most one solution.
 %In elastodynamics, this condition implies the reality of wave speeds \cite{ball10}.
In vector space,   the  LH condition is equivalent to Legendre condition $\nabla^2 \WW(\bgamma) \succeq 0  \;\; \forall \bgamma \in \real^n$.

Clearly, the LH condition is only a sufficient condition  for local minimizer of the variational problem $(\calP)$.
 In order to identify ellipticity, one must to check LH condition
for all local solutions, which is impossible for general fully nonlinear problems.
%By the canonical duality theory we can show that the LH condition  is   not necessary for  ellipticity
 %of  general  nonconvex systems (see Section 3).
 Also, the traditional  ellipticity  definition  depends only on the stored energy $\WW(\bF)$
regardless of the linear term in $\PP(\bF)$.
This definition works only for convex systems since the linear term can't change the convexity of $\PP(\bF)$.
But this is not true for  nonconvex  systems.
To see this, let us consider the St. Venant-Kirchhoff material
\eb
W(\bF) = \half \bE : {\bf H} : \bE, \;\; \bE = \half [(\bF)^T(\bF) - \bI ],
\ee
where $\bI$ is  a unit tensor in $\real^{3 \times 3}$. Clearly, this function is not even rank-one convex.
A special case of this model is the well-known double-well potential
$\WW(\bgamma) = \half ( \half |\bgamma|^2- 1)^2$.
In this case, if we let $\xi = \Lambda(\bgamma) = \half   |\bgamma|^2 - 1$ be  an objective measure, we have the
canonical function $\barW(\xi) = \half \xi^2 $. In this case, the canonical dual algebraic equation
(\ref{cda}) is  a cubic  equation (see \cite{gao-dual00})
$2 \zeta^2 (\zeta +1) = \tau^2$, which has at most three real solutions $\{\zeta_k(\bx) \}$ at each $\bx \in \calB$
satisfying $\zeta_1 \ge 0 \ge \zeta_2\ge \zeta_3$.
It was proved in \cite{gao-dual00} (Theorem 3.4.4, page 133)
that for a given force $\bt(\bx)$,
if  $\tau^2(\bx) > 8/27 \;\; \forall \bx \in \calB \subset \real$, then $(BVP)_g$ has only one solution on $\calB$.
If $\tau^2(\bx) < 8 /27 \;\; \forall \bx \in \calB_s  \subset \calB$, then
 $(BVP)_g$
has   three solutions $\{\chi_k(\bx)\}$ at each $\bx\in \calB_s$, i.e. $\Pi(\chi)$ is nonconvex on $\calB_s$.
It was shown by Gao and Ogden that these solutions are nonsmooth if  $\tau(\bx)$ changes its sign on $\calB_s$
 \cite{gao-ogden-qjmam}.

Analytical solutions for general 3-D finite deformation problem $(\calP)$ were first proposed  by Gao in 1998-1999 \cite{gao-ima98,gao-mecc99}.
It is proved recently \cite{gao-haj} that for St Venant-Kirchhoff material, the problem $(\calP)$ could have 24 critical  solutions at
each material point $\bx \in \calB$, but only one global minimizer.
The solution is unique if the external force is sufficiently bigger.

\begin{figure}[h]
\begin{center}
 \includegraphics[width=12cm,height=4.5cm]{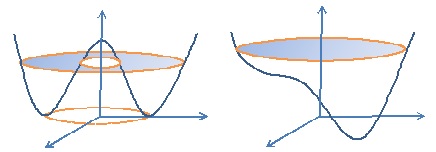}
 \caption{Graphs and level sets of $P(\bx)$ with $ {\btau}= 0$ (left) and $ {\btau} \neq 0$ (right) }
 \label{1-dw} \vspace{-.4cm}
 \end{center}
 \end{figure}

The geometrical explanation for ellipticity and Theorem \ref{thm-1} is illustrated by
 Fig. \ref{1-dw}, which shows that
 the nonconvex function
$\PP(\bgamma) = \half ( \half  |\bgamma|^2- 1)^2 - \bgamma^T \btau$ depends sensitively  on the
external force $\btau \in \real^2$.
If $|\btau |$ is bigger enough,  $\PP(\bgamma) $
  has only one  minimizer and its
  level set  is an ellipse (Fig. \ref{1-dw} (b)).
  Otherwise,  $P(\bgamma)$ has multiple local minimizers and its level set is not an ellipse.
  For $\btau = 0$, it is  well-known  Mexican-hat in theoretical physics (Fig. \ref{1-dw} (a)). \vspace{-.3cm}

  Fig. \ref{1-dw}   shows    that although 
  $P(\bgamma)$ has only one global minimizer for certain given ${\btau}$, the function is still nonconvex.
  Such a function is called quasiconvex in the context of global optimization.
   In order to distinguish this type of functions with  Morry's    quasiconvexity in nonconvex analysis,
   a generalized definition on a tensor space $\calF_a \subset \real^{m\times n}$ could be convenient.

  \begin{definition}[G-Quasiconvexity]
  A function $\PP:\calF_a \subset \real^{m\times n} \rightarrow \real $  is called G-quasiconvex
  if its domain $\calF_a$ is convex and
  \eb
  \PP( \theta \bF + (1- \theta) \bT) \le \max \{ \PP(\bF) , \PP(\bT) \} \;\;
  \forall \bF, \; \bT  \in \calF_a, \;\; \forall \theta \in [0,1].
  \ee
  It is called strictly G-quasiconvex if the inequality holds strictly.
\end{definition}
 For a given function $\PP:\calF_a  \rightarrow \real $, its {\em level set } and  {\em sub-level set} of height $\alpha \in \real$
 are defined, respectively, as the following
  \eb
 {\cal L}_\alpha(\PP) = \{ \bF \in \calF_a\; | \;\; \PP(\bF) =  \alpha \}, \; \;    {\cal L}^\flat_\alpha(\PP) = \{ \bF \in \calF_a\; | \;\; \PP(\bF) \le \alpha \}, \; \;   \alpha \in \real.
  \ee
Moreover, we may need  a  generalize   ellipticity  definition  for nonconvex systems.
\begin{definition}[G-Ellipticity]
 For a given function $\PP:\calF_a \rightarrow \real$ and $\alp \in \real$, its
 level set ${\cal L}_\alp(\PP)$ is said to be a G-ellipse if it is a closed, simply connected set.
For a given $\bt $ such that $\bT \in \calT_a$, the $(BVP)$ is said to be $G$-elliptic if the total potential function
$\PP(\bF)$ is G-quasiconvex on $\calF_a$. $(BVP)$ is strongly elliptic if $\PP(\bF)$ is strictly G-quasiconvex.
\end{definition}

  Clearly, we have the following statements:
 \begin{eqnarray}
 \mbox{  $\PP(\bF)  $  is  G-quasiconvex  $\Leftrightarrow $  ${\cal L}^\flat_\alpha(\PP) $  is  convex
 $\Leftrightarrow $  ${\cal L}_\alpha(\PP) $  is  G-ellipse   $\forall \alpha \in \real$  .} \\ 
 \mbox{  $\PP(\bF)  $  convex $\Rightarrow $  rank-one convex $\Rightarrow $  G-quasiconvex  $\Rightarrow $  $(BVP)$ is G-elliptic} .
 \end{eqnarray} 

This statement shows a fact  in nonconvex systems, i.e.
  the number of solutions to a nonlinear equation depends not only on the stored energy, but also (mainly)  on  the external force field.
The nonlinear partial differential equation  in $(BVP)$
is elliptic only if it is  G-elliptic. $(BVP)$ has at most one  solution  if the integrand $\PP(\bF)$ in the total potential $\Pi(\bchi)$
is strictly  G-quasiconvex on $\calF_a$.

In global optimization,   the most simple
 quadratic integer programming problem
\[
(\calP)_i : \;\;\; \min \left\{ \Pi(\bx) = \half \bx^T \bQ \bx -  \bx^T {\bf t} \;\; | \;\; \bx \in \{ 0, 1\}^n \subset \real^n \right\}
\]
could have up to $2^n$ local minimizers  due to the indefinite matrix $\bQ$ and the integer constraint.
Such a nonconvex discrete optimization problem is considered as NP-hard in computer science.
However, by using canonical transformation $\bxi  = \{ x_i(x_i -1)\}$,    the canonical dual of this discrete problem  
is  a concave maximization over a convex set in continuous space. %, which can be solved easily if $\calS^+_a \neq \emptyset$. 
It was proved in \cite{gao-cace09} that as long as  the
source term ${\bf t} = \{t_i\} $ is bigger enough,  $\calS^+_a \neq \emptyset$ and $(\calP)_i$ is not NP-hard.
 The decision variable is simply $\{ x_i\} = \{0  \mbox{ if }  t_i  < 0 , \; 1 \; \mbox{ if } t_i > 0\}$ (Theorem 8, \cite{gao-cace09}).

\section{Anti-plane Shear Deformation Problems}
Now let us consider a special case that  the  homogeneous  elastic body
$\calB\subset \real^3$ is a cylinder
with generators parallel to the $\be_3$ axis and
with cross section a sufficiently nice region $\Oo \subset \real^2$ in the
$\be_1 \times \be_2$ plane.
The so-called anti-plane shear  deformation is defined by (see Knowles 1976, \cite{knowles})
\begin{equation}\label{defor}
\bchi (\bx) = \left\{ \lambda^{-\half} x_1, \;\; \lambda^{-\half} x_2, \;\;
 \lambda x_3  + \uu(\xx_1, \xx_2) \right\} : \Oo \rightarrow \real^3,
\end{equation}
where $ (\xx_1, \xx_2, \xx_3 )$ are cylindrical  coordinates in the
reference configuration  $\calB$ relative to a
cylindrical  basis $\{\be_i\},\; i = 1,2,3$, the parameter
$\lambda $ is a positive constant, and $\uu:\Oo \rightarrow \real$ is
 the amount of shear (locally a simple shear) in the planes normal to $\be_3$.
 On $\Gu \subset \partial \Oo$, the homogenous boundary condition is given
$
 \uu(x_\alp)= 0 \;\; \forall x_\alp \in \Gu, \;\; \alp = 1,2 .$
On the remaining boundary $\Gt = \partial \Oo  \cap \Gu$, the cylinder is subjected to the  shear force
\[
{\bf t} (\bx) =  t(\bx)  \be_3 \;\; \forall \bx \in \Gt  ,
\]
where   $ t: \Gt \rightarrow \real $ is a prescribed function.
According to Knowles, the
deformation (\ref{defor})   may be thought of as one in which the body first undergoes
an axial elongation (or contraction) of stretch ratio $\lambda$ (regarded as given), and
is then subjected to an anti-plane shear with out-of-plane displacement $\uu$.
For this anti-plane shear deformation we have
\begin{equation}\label{1}
\mathbf{F}= \nabla \bchi =  \left(\begin{array}{ccc}
 \lambda^{-\half} & 0 & 0 \\
0 &  \lambda^{-\half}& 0 \\
\uu_{,1} & \uu_{,2} & \lambda  \end{array} \right), \;\; \;\;
\mathbf{C}=\mathbf{F}^{\rm
T} \bF =
\left( \begin{array}{ccc}
  \lambda^{-1} + \uu_{,1}^2  & \uu_{,1} \uu_{,2} & \lambda \uu_{,1}\\
\uu_{,1} \uu_{,2}&  \lambda^{-1} + \uu_{,2}^2 & \lambda \uu_{,2} \\
 \lambda\uu_{,1} &  \lambda\uu_{,2} &  \lambda^{2}
\end{array} \right) ,
\end{equation}
where  $\uu_{,\alpha}$ represents $\partial \uu/\partial x_\alpha$ for $\alpha= 1,2$.
By the notation $|\nabla \uu|^2 = \uu_{,1}^2 + \uu_{,2}^2$, we have
\begin{equation}
I_1(\bC)= \lambda_1  + | \nabla \uu |^2, \;\;
 I_2 (\bC)= \lambda_2  +  \lambda^{-1} | \nabla \uu |^2 ,  \;\;
  I_3 (\bC)  \equiv 1  , \la{Ii}
\end{equation}
where $\lambda_1 = \lambda^2 + 2 \lambda^{-1}, \; \lambda_2 = \lambda^{-2} + 2 \lambda$.
Particularly,
\eb
I_1(\bC)=
 I_2 (\bC)=  3 +  | \nabla \uu |^2  \;\; \mbox{ if } \lambda =1.
 \ee
 \begin{Lemma}\label{lemneo}
For any given  $\lambda> 0$, the homogenous hyper-elasticity  for   general anti-plane shear deformation  must be governed by a generalized neo-Hookean model, i.e.
$\WW(\bF) = \barW (I_1)$.
 Particularly,  the  Mooney-Rivlin model is identical to the neo-Hookean model subjected to a constant, i.e.
$
\WW(\bF) = A (I_1 - 3) + B
$
with  $B= 0$ if $\lambda = 1$.
\end{Lemma}

The proof  is elementary, i.e. by  the fact that  $I_2  = \lambda^{-1} I_1 + a$, $a =  \lambda_2 - \lambda^{-1} \lambda_1 $, we have
\[
\WW(\bF) =  \bar{W}(I_1, I_2) =  \bar{W}(I_1, \lambda^{-1} I_1 + a ) = \barW(I_1) , \;\;\forall \lambda > 0.
\]

Moreover, let  $A = c_1 + c_2 \lambda^{-1}, \;\;
B = c_2 (3 \lambda^{-1} - 3 +  \lambda_2 - \lambda_1 \lambda^{-1})$, we have the  Mooney-Rivlin model $\WW(\bF) = c_1 (I_1 -3) + c_2 (I_2- 3)=  A (I_1 - 3) + B $.\\

The fact   $\det \bF \equiv  1 $  shows that the anti-plane shear state  (\ref{defor})
  is an isochoric   deformation.
Therefore,
 the kinetically admissible displacement space $\calX_c$   can be simply replaced by a convex set
\eb
\calU_c = \{ \uu(x_1,  x_2)\in \calW^{1,1}(\Oo; \real)| \;\; \uu(x_\alp)= 0 \;\; \forall x_\alp \in \Gu \}. \label{eq-uc}
\ee
Thus, in terms of $\bgamma = \nabla \uu$,  $\xi =   \Lambda( \bgamma) = |\bgamma|^2  + \lambda_1$, and  $\WW(\bF(\bgamma)) = \barW(\Lambda(\bgamma)) $, for any given
 \[
 \btau \in  \calT_a =  \{ \btau \in C^1[\Oo; \real^2] | \;\; \nabla \cdot \btau = 0  \;\; \mbox{ in } \Oo,
\;\; \bn \cdot \btau = t \;\; \mbox{ on } \Gt \}
\]
Problem $(\calP)_g$ for the general anti-plane shear deformation problem has the following form
\eb
(\calP)_s: \;\; \min \left\{ \Pi(\uu) = \int_\Oo \PP(\nabla\uu) \dO \;\;| \;\; \uu \in \calU_c \right\}, \;\;
\PP(\bgamma) = \barW(\Lambda(\bgamma)) - \bgamma^T \btau
\ee

%Let $\calU_b  $ be a smooth subset of $\calU_c$ defined by
%\eb \calU_b = \{ \uu \in \calU_c | \; \uu \in C^2(\Oo; \real), \;\;\barW(\xi) \in C^2[\calE_a;\real]  \}
%\ee
Under certain regularity conditions, the associated  mixed boundary value problem is
\eb\label{eq-eqs}
(BVP)_s: \;\;    \left\{ \begin{array}{l}
\nabla \cdot \left( 2  \zeta   \nabla  \uu  \right)  = 0 \;\; \mbox{ in } \Oo, \\
\bn \cdot \left( 2  \zeta   \nabla  \uu  \right)= t \;\; \mbox{ on } \Gt, \;\; \uu = 0 \;\; \mbox{ on } \Gu
\end{array} \right.
\ee
where  $ \bn =  \{ n_\alpha\} \in \real^2$ is a unit vector  norm to $\partial \Oo$, and
$ \zeta  = \nabla  \barW(\xi), \;\;  \xi = \lambda_1 + |\nabla \uu|$.

If $\Gu = \partial \Oo$, then $(BVP)_s$ is a Dirichlet boundary value problem, which has only trivial solution due to zero input. For Neumann boundary value problem $\Gt = \partial \Oo$, the external force field must be such that
\[
\int_\Gt t(\bx) \dG  = 0
\]
for overall force equilibrium. In this case, if $\barbchi$ is a solution to  $(BVP)_s$, then $\bchi = \barbchi + {\bf c}$ is also
a solution for any vector ${\bf c} \in \real^3$ since the cylinder is not fixed.

By the fact that the only unknown $\uu$ is a scalar-valued function,
 anti-plane shear deformations are one of the simplest classes of deformations that solids can undergo \cite{horgan}.
Indeed, if $\barW(\xi)$ is a canonical function
and for any given
$ \btau \in \calT_a$
 such that $\tau= |\btau|$,
 the canonical dual problem has a very simple form
  \eb
 (\calP^d)_s: \;\;\;\; \sta \left\{  \Pi^d(\vsig) =   \int_{\Oo}  \left[
 \lambda_1 \zeta  - \barW^*(\vsig) - \frac{1}{4} \vsig^{-1}  \tau^2  \right] \dO \; | \;\; \vsig \in \calS_a \right\} .
  \ee
Since  $\Lambda(\uu) = |\nabla \uu|^2 + \lambda_1$, the canonical dual equation (\ref{cda}) for this problem is
\eb
4 \zeta^2 [ \nabla \VV^*(\zeta) - \lambda_1] = \tau^2. \label{cda1}
\ee
\begin{thm}\label{thm-equiv}
For any given pre-stretch $\lambda > 0$ and  non-trivial shear force $ t (\bx)   \neq 0 $ on $ \Gt$ such that $\calT_a \neq \emptyset$, the canonical dual problem $(\calP^d)_s$  has at least one non-trivial solution   $\vsig_k$  and
   \eb\label{eq-asolu}
   \uu_k (\bx) = \half \int_{\bx_0}^\bx  \vsig_k^{-1}  \btau \cdot \mbox{d}\bx
   \ee
   along any path from $\bx_0 \in \Gu$ to $\bx \in \Oo$ is a critical point of $\Pi(\uu)$ and
   $\Pi(\uu_k) = \Pi^d( \vsig_k) $.
 \end{thm}
 {\bf Proof.} By the fact that the  nontrivial shear force $\bt$  leads to $\tau > 0$.
  The  equation  (\ref{cda1})   has  at least one nontrivial solution
  $\vsig_k \neq 0$.
  By the pure complementary energy principle, the equation (\ref{eq-asolu}) gives a nontrivial solution to
  the anti-plane shear deformation. \hfill $\Box$\\

Note that   $\bF$ is an affine function of   $\bgamma = \nabla \uu \in \real^2$,
it is also  mathematically equivalent  to assume
  the existence of  a real-valued function $\hatW:\real^2 \rightarrow \real$ such that
\eb
 \WW(\bF(  \uu)) = \barW(I_1(\nabla \uu))= \hatW(\bgamma( \uu) )\;\; \forall \uu \in \calU_c
 \ee
 holds  for general  anti-plane shear deformation problems  without any additional constitutive constraints.
   In this case, by choosing $\xi = \Lambda(\nabla \uu) = |\nabla \uu|^2$ and
  the canonical transformation
  $\VV(\xi(\bgamma)) =  \hatW(\bgamma  )  $,
 equivalent results for  complete set of  solutions have been obtained for both convex and nonconvex
  anti-plane shear deformation problems \cite{gao-cmt15}.

 Clearly, the anti-plane shear problem $(BVP)_s$ is linear only if   $\barW(I_1) = A(I_1 -3)$ for a given constant $A> 0$, i.e. the   neo-Hookean model.
For nonlinear elasticity, the  problem $(\calP)_s$ could have multiple critical solutions $\{\uu_k(\bx)\}$ at each $\bx \in \Oo_s \subseteq \Oo$. As long as $\Oo_s \neq \varnothing$, the boundary value problem $(BVP)_s$ should
 have infinitely  many  solutions (see \cite{gao-ogden-qjmam}).
 Therefore, it is impossible to use Legendre condition to identify global minimal solution.
 Theorem \ref{thm-1} shows that the  Legendre condition is only  necessary
  but not sufficient condition for global optimality.
 %\eb \nabla^2 \WW(\bgamma_k) = 2 \vsig_k \bI\otimes \bI   + \vsig_k^{-2}  h(\xi_k)  \btau  \otimes \btau  ,\ee
  The sufficient condition is simply
\eb\label{ellip}
 \vsig_k  \in \bar{\calS}^+_a   \;\; \Leftrightarrow \;\; G_{ap} = \int_\Oo \vsig_k |\nabla \uu|^2 \dO \ge 0 \;\; \forall \uu\in \calU_c, \;\;  \vsig_k \in \bar{\calS}^+_a,
\ee
which  was first proposed   in 1992 \cite{gao-zamp92}.
If all solutions $\{ \zeta_k \} \in \bar{\calS}^+_a$,
Problem   $(\calP)_s$  is G-quasiconvex,  which has unique solution if $\{ \zeta_k \} \in \calS^+_a$.
   Application  of Theorem \ref{thm-equiv}  has  been illustrated for both convex and nonconvex problems given recently in \cite{gao-cmt15}.

\section{Remarks on Knowles' over determined problem}
Now let us revisit Knowles' work in 1976 \cite{knowles}.
 Instead of the  minimal  potential variational problem  $(\calP)$, Knowles started from the
 strong form of  $(\calP)$, i.e. $\div \bsig = 0$ in
  the boundary value problem $(BVP)_p$  given in (\ref{eq-geqi}) with
general constitutive law  for incompressible materials
\eb
  \bsig  = \frac{\partial  \bar{W}(I_1,I _2)}{\partial \bF}   - p \bF^{-T}.
\ee
For  the same anti-plane shear deformation problem  (\ref{defor}),
 he ended up with  three equilibrium equations (i.e.  equations  (2.19) and (2.20) in \cite{knowles})\footnote{There is a mistake in \cite{knowles}, i.e.  $\bar{W}_1$ in  Knowles' equation (2.19) should be $\bar{W}_2$}:
 \eb\label{2.19}
 q_{,\alp} + \left( 2 \bar{W}_2  \uu_{,\alp} \uu_{,\beta} \right)_{,\beta} - p_{,3} \uu_{,\alp} = 0,
 \ee
 \eb\label{2.20}
 \left[ 2 (\bar{W}_1 + \lambda^{-1} \bar{W}_2) \uu_{,\beta} \right]_{,\beta} - \lambda^{-1} p_{,3} = 0,
 \ee
 where
 $\bar{W}_\alp = \partial \bar{W}/\partial I_\alp = \zeta_\alp, \;\; \alp, \beta = 1,2$ and
$
 q = \lambda p - 2  \bar{W}_1 - 2 (\lambda^2 + \lambda^{-1} + |\nabla \uu|^2) \bar{W}_2.$

The first two  equations in  (\ref{2.19}) are corresponding to the general equilibrium equation $\sigma_{ij,j} = 0$ in $\be_1$ and $\be_2$ directions;
while the third one  (\ref{2.20}) is  in $\be_3$ direction.
Knowles  indicated (Equation (2.22) in  \cite{knowles})  that the hydrostatic pressure $p$ is linear in $x_3$, i.e.
\eb
p= c x_3 + \bar{p}(x_1,x_2) \label{2.22}
\ee
 where $c$ is a constant. Saccomandi  emphasized recently    that $p$ is the Lagrange multiple   associated with the incompressibility constraint, which must be in the form of (\ref{2.22})  and $c \neq 0 $   for general incompressible material \cite{sacc15}.

Clearly, for a given strain energy $\WW(\bF) = \bar{W}(I_1, I_2)$, the governing equations obtained by Knowles
constitute an over-determined system in general, i.e.
  two unknowns $(\uu, p)$ but  three equations.
  In order to solve this over determined problem, Knowles believed that  the
   stored energy $\bar{W}(I_1,I_2)$  should have some restrictions and he proved the following theorem. \vspace{.2cm}\\
  {\bf Theorem (Knowles, 1976 \cite{knowles})} {\em If
  the stored energy $\bar{W}(I_1, I_2)$ is such that   the  ellipticity condition (i.e. the equation (3.5) in \cite{knowles})
  \eb\label{3.5}
  \frac{d [2 R (\bar{W}_1 + \lambda^{-1} \bar{W}_2)]}{dR} > 0 \;\; \;  \forall R \ge 0, \;\; \lambda > 0
  \ee
 holds,  then
  the associated incompressible elastic material    admits nontrivial states of
anti-plane shear   for a given pre-stretch $\lambda$ if and only if $\bar{W}(I_1, I_2)$ also satisfies
the following  constitutive constraint    (i.e. equation (3.22) in \cite{knowles})
  \eb\label{3.22}
  b \bar{W}_1 + (b \lambda^{-1} - 1) \bar{W}_2  = 0  ,
  \ee
 for some constant   $b$,  for all values of $I_1, I_2$ such that
$  I_1 = \lambda_1  + R^2 , \;\; I_2 = \lambda_2  + \lambda^{-1} R^2, \;\;  R = |\bgamma| . $}\\

First, by Lemma \ref{lemneo} we know that
 $\WW(\bF) = \bar{W}(I_1, I_2) = \barW(I_1)$ hold for any given anti-plane shear deformation.
There is no need to have both $I_1, I_2$ as variables.
Therefore, the following  trivial  result   shows immediately that
 Knowles' condition (\ref{3.22}) is not a constitutive constraint.\vspace{-.2cm}
 \begin{Lemma}\label{lem31}
For any given stored energy $\WW(\bF) = \bar{W}(I_1,I_2)$ such that
$
 I_1 = \lambda_1 + | \nabla\uu|^2, \;\;
I_2 = \lambda_2  +  \lambda^{-1} | \nabla \uu |^2   $,
  Knowles' constitutive  condition (\ref{3.22}) is automatically satisfied for $b = \half \lambda$.
\end{Lemma}

The proof of this statement  is  elementary: by chain rule and  $I_1 = \lambda I_2 + \lambda_1 - \lambda \lambda_2 $, we  have
immediately
\[
\bar{W}_2 = \frac{\partial \bar{W}}{\partial I_1} \frac{\partial I_1}{\partial I_2}
= \lambda \bar{W}_1
 \Rightarrow \;\;
 b \bar{W}_1 + (b \lambda^{-1} - 1) \bar{W}_2 = (2 b - \lambda) \bar{W}_1 = 0 \;\; \forall b = \half \lambda .
 \]

 To check if the Lagrange multiplier   $p=p(x_1,x_2,x_3)$ must be in  Knowles' formula (\ref{2.22}),
we use mathematical theory of Lagrange duality.
 For any given real-valued function $\phi(\bx) \in L^q(\calB)$, the Lagrange multiplier $p$ for the equality constraint
  $\phi(\bx) = 0$ must
 be in the dual space $L^{q'}(\calB)$ such that   $1/q + 1/q' = 1$. Since
   $ \bF $ depends only on $(x_1,x_2)\in \Oo$, the constraint  $\phi(\bx) = \det \bF(\uu) - 1$ is defined on $\Oo \subset \calB$, its Lagrange multiplier $p(\bx)$ must be defined on $\Oo$.
 Indeed,  by simple calculation for the form  (\ref{2.22})
 \[
 \int_\calB   \phi(x_1,x_2) p(x_1,x_2,x_3) \dB =  \int_\Oo \phi(x_1,x_2)  [ \int  p(x_1,x_2,x_3) d x_3 ]  \dO
  \]
    one can easily find that the Lagrange multiplier is independent of $x_3$. Thus,  we must have  $c \equiv 0$  and $p = p(x_1,x_2)$
for any anti-plane shear deformation.
For this reason and   $\bar{W}_2 = \lambda \bar{W}_1$,  $\vsig = \nabla \barW(\xi) = \bar{W}_1$,  the  equation  (\ref{2.20}) (i.e. (2.20) in \cite{knowles}) is identical to the equation in $(BVP)_s$:
\eb\label{3.4}
 \left[ 2 (\bar{W}_1 + \lambda^{-1} \bar{W}_2) \uu_{,\beta} \right]_{,\beta}  = 0 \;\; \Leftrightarrow \;\;
 \nabla \cdot  [\zeta \nabla \uu ] = 0 \;\; \mbox{ in } \Oo .
\ee

Now we need to check the other two equilibrium equations in  Knowles' over-determined system.
Instead of the local analysis, we use the well-known {\em virtual work principle}
\eb
\int_{\calB} \tr (\bsig \cdot \delta \bF(\bchi) ) d\calB = \int_{S_t} \bt \cdot \delta\bchi, \;\;  \;\;
\forall \bchi \in \calX_c
\ee
which holds for any given deformation problem regardless of constitutive laws.
For smooth deformation $\bchi$  and sufficiently regular  $\calB$ and $\partial \calB$, we have the following
   strong  complementarity conditions %The strong form of this principle is
 \eb
 (\delta \bchi) \cdot (\div \bsig) = 0   \;\; \mbox{ in } \calB, \;\;   (\delta \bchi) \cdot  \bsig \cdot \bN =  ( \delta \bchi) \cdot \bt  \;\; \mbox{ on } S_t
 \ee
The fact that the anti-plane shear deformation (\ref{defor}) has no displacements $ \{\uu_i\}$ in $\be_1$ and $\be_2$ directions,
 i.e. $  \delta \chi_\alp  \equiv 0  \;\; \forall  \alp = 1,2 \;\; a.e.$ in $\Oo$,
  the vector  $\div \bsig$ is  not  necessarily to be  zero in these directions.
This shows that   the additional two equilibrium equations (\ref{2.19}), i.e.
 (2.19) in the  paper \cite{knowles},  can't be obtained from the virtual work principle.
 By the fact that the boundary value problem $(BVP)_s$ is well-determined by the equation  (\ref{3.4}),
 these two extra equations are useless for the problem considered.

To  understand the ``function" of  the  hydrostatic pressure $p(\bx)$  in Knowles' over-determined problems
for either compressible or incompressible materials,
we use the    KKT complementarity condition  in (\ref{eq-compf}), i.e.
$p(\det \bF - 1) = 0$.
As we know  that the anti-plane shear state  is a volume preserving deformation, the
equality $ \det \bF(\uu)  \equiv 1$ is trivially  satisfied all most every where in $\Oo$ for any materials.
Thus, we must have  $p(\bx) \neq 0 \;\; a.e. $ in $\Oo$, i.e.  the only function of this arbitrary non zero
 parameter is to balance the extra two equations (\ref{2.19}),
which can't be obtained by the virtual work principle.
 This shows that the governing equations obtained by the minimum total potential principle are always compatible.

Finally,  let us exam  the ellipticity condition  in Knowles's  theorem.
On page 407 of   \cite{knowles}, Knowles indicated:
 the condition  (\ref{3.5})  ``guarantees that (\ref{3.4}) is elliptic at every solution $\uu$  and at every point in $\Oo$".
  The following theorem is  important in nonlinear analysis.
  \begin{thm}\label{thm3}
The ellipticity condition (\ref{3.5}) is neither  necessary nor sufficient
 for the nonlinear PDE (\ref{3.4}) to admit  nontrivial states of anti-plane shear.
 $(BVP)_s$ has at least one solution   only if $t(\bx) \neq 0$ on $\Gt$ such that $\calT_a \neq \emptyset$.

   For any given convex function $\bar{W}(I_1,I_2)$  and the external force $t(\bx)  \neq 0$ on $\Gt$,
  the equation (\ref{3.4}) is strongly  G-elliptic if
  \eb
 \zeta_1   >  0,   \;\; \zeta_1 = \bar{W}_1(I_1,I_2)  \label{eq-s}
  \ee
  for  every  solution $ \zeta_1 $ of (\ref{cda1}).
  \end{thm}
  {\bf Proof.} Let $\bxi = \{ I_1, I_2\}$. By using chain rule for
  $\hatW(\bgamma) = \bar{W}(\bxi(\bgamma))$% = \VV(I_1)$
\[
\nabla \hatW(\bgamma) = \nabla_{\bgamma} \bar{W}(\bxi(\bgamma)) = 2 \bgamma (\bar{W}_1 + \lambda^{-1} \bar{W}_2),
\]
thus, Knowles' ellipticity  condition (\ref{3.5})   is actually a special case of the strong Legendre condition
 $\nabla^2 \hatW(\bgamma) \succ 0$,
 which can   only guarantee the convexity of  $\WW (\bF) = \hatW(\bgamma)$, i.e. under this condition, the
 $(BVP)_s$ has at most one solution.
 Clarly,  $(BVP)_s$  has a trivial solution if $t(\bx) = 0$ on $\Gt$. Therefore, Knowles' ellipticity  condition (\ref{3.5})
 is not sufficient to admit a nontrivial solution.

 By the canonical duality theory we know that for nonconvex stored energy $\WW(\bF) = \hatW(\bgamma)$, the $(BVP)_s$ has multiple
 nontrivial solutions if $t(\bx) \neq 0$ on $\Gt$ such that $\calT_a \neq \emptyset$.
 Therefore, Knowles' ellipticity  condition (\ref{3.5})
 is also not necessary  to admit a nontrivial solution.
 
 By simple calculation  for (\ref{3.5}), we have
 \eb \label{eq-caco}
 2 (\bar{W}_1 + \lambda^{-1} \bar{W}_2) + 4 R^2(\bar{W}_{11} + 2 \lambda^{-1} \bar{W}_{12} + \lambda^{-2} \bar{W}_{22} ) > 0 ,
 \ee
 which is a strong case  for (\ref{eq-hatw}), where $\bar{W}_{\alp\beta} = \partial^2\bar{W}/\partial I_\alp \partial I_\beta$.
If  the canonical function $\bar{W}(I_1,I_2)$ is convex in $\bxi = \{I_1, I_2\}$, we have
 \eb\label{barwc}
  \bar{W}_{11} + 2 \lambda^{-1} \bar{W}_{12} + \lambda^{-2} \bar{W}_{22}   \ge  0 \;\; \forall \{I_1, I_2\} \in \real^2, \;\; \lambda > 0.
  \ee
  By the facts that
  $\zeta_2 = \bar{W}_2 = \lambda \bar{W}_1 = \lambda \zeta_1$ and $\zeta_1 = \nabla \VV(I_1) = \bar{W}_1$,  we know that  the condition (\ref{eq-caco}) holds as long as
 \[
 2 (\bar{W}_1 + \lambda^{-1} \bar{W}_2) = 4 \zeta_1  >  0  .
 \]
  Thus, by Theorem \ref{thm-1}     we know that the function $\PP(\bgamma)$ is strictly G-quasiconvex and (\ref{3.4}) is strongly G-elliptic. In this case, $(BVP)_s$ has at most one solution.
  \hfill $\Box$\\

Combining Theorems  \ref{thm-equiv},    \ref{thm3} and Lemma \ref{lem31} we know that
 Knowles' constitutive constraints   (\ref{3.5}) and (\ref{3.22}) are neither  necessary
 nor sufficient for  the existence of  nontrivial states of anti-plane shear.
 Actually, this ellipticity condition even disallows many possible nontrivial local  solutions in  nonconvex problems.
  Indeed, it was shown in \cite{gao-cmt15,gao-ogden-zamp} that for any given nonconvex stored
 energy $\WW(\bF(\bgamma)) = \bar{W}(I_1(\bgamma),I_2(\bgamma)) = \hatW(\bgamma)$ and
 nontrivial external force  $t(\bx) \neq 0$, the minimum potential variational
problem  $(\calP)_s  $ has at least one solution  $\{\uu_k\} $ in Banach space $\calU_c$,
which can be obtained analytically by the canonical duality theory. If $t(\bx)$ is very small, the solution may not unique,
the one such that $\vsig(\uu_k) \in \calS^+_a$ is a global minimal solution.
Both global and local minimum solutions could be nonsmooth if $\btau(\bx)$ changes its sign in $\Oo$.
While Knowles' over-determined system admits only a unique smooth solution in $C^2$ due to the additional    ellipticity restriction on
$\bar{W}$. Therefore, Knowles' over-determined system  is a very special case of  the variational problem $(\calP)_s$.

\section{Conclusions}
In summary,  the following conclusions can be   obtained. % \cite{knowles,knowles1}:
\begin{verse}
1.  The pure complementary energy principle and canonical duality-triality  theory developed in \cite{gao-dual00} are useful
 for
 solving general nonlinear boundary value problems in nonlinear elasticity. \\

 2. The ellipticity condition for fully nonlinear boundary value problems
 in finite deformation theory depends not only on the stored energy function,
 but also on the external force field.\\

 3.   %The Legendre  condition is only a necessary ellipticity condition for convex systems.
 The   triality theory provides a sufficient condition to identify  both global and local extremum solutions for nonconvex problems.\\

4. General anti-plane shear deformation problems must be governed by the generalized neo-Hookean model. \\

5. Unless the KKT theory is wrong, the incompressibility  is not a variational constraint for any anti-plane shear deformation problem, the
  pseudo-Lagrange multiplier $p$ depends only on $(x_1,x_2)$, which  is not a  variable for the problem.  \\

6. Unless the virtual work principle is wrong, there is only one equilibrium equation for general  anti-plane shear deformation problems.
The two  extra equations  % (\ref{2.19})
in  Knowles' over-determined system  are not required.\\

7. Unless the  minimum potential variational principle    is wrong, % for general finite elasticity,
the constitutive conditions  required by  Knowles'  Theorems   in \cite{knowles,knowles1} are neither   necessary nor sufficient
 for general  homogeneous  materials to admit nontrivial states of anti-plane shear.

%5. Based on the variational principle in  finite elasticity, Gao's  analytical solutions presented in
 %  \cite{gao-cmt15} are correct for  general anti-plane shear deformation problems, which could be nonsmooth in Banach space for certain  external force field. \\

%6.   G.  Saccomandi's criticisms in  \cite{sacc15} are incorrect, which are  based on Knowles' overdetermined system  and
 % this  system allows  only unique smooth solution in $C^2$ space to special anti-plane shear problems.
  \end{verse}

The first three  conclusions are naturally included in the canonical duality-triality theory developed by the author and his co-workers during the last 25 years \cite{gao-dual00}.
Extensive applications have been given in multidisciplinary fields of biology, chaotic dynamics, computational mechanics, information theory, phase transitions, post-buckling,  operations research, industrial and systems engineering, etc.  (see  recent review article \cite{gao-bridge}).

The last four conclusions are obtained recently when the author got involved in the discussions with colleagues
on anti-plane shear deformation problems.
%Results  presented  in \cite{knowles} were  directly applied for a special case $\lambda = 1$ in \cite{knowles1}.
As highly cited papers  \cite{knowles,knowles1},  Knowles' over-determined system has been extensively
 applied to many anti-plane shear deformation problems   in literature, see recent papers \cite{pucci-r-s14,
 pucci-sacc,pucci-sacc1,sacc15}.
This  is the  motivation for  this paper.
%say recent papers  {\em et al}
%}, especially in the recent paper by G.  Saccomandi \cite{
%inequality (\ref{3.5}) (i.e.  (3.5) in \cite{knowles})
% can't guarantee the equilibrium equation (\ref{3.4}) (i.e.   (3.4) in \cite{knowles})  is  elliptic
%  unless the stored energy $\WW(\bF(\bgamma))$ is convex in $\bgamma$.
 %  For general nonconvex $\WW(\bF)$, the ellipticity condition is given by (\ref{ellip}).
    %if $\bar{W}_{11} + 2 \lambda^{-1} \barW_{12} + \lambda^{-2}\barW_{22} \ge 0$.

 %The last conclusion is the only motivation for writing this paper.
% But  these are not the motivation for the author to  write this paper
% since his  attention  is  only focused on people's ideas and methodologies  instead of mistakes.
% The only reason for writing  this paper is that the author's recent work on the anti-plane

\subsection*{Acknowledgements}
Insightful discussions with  Professor David Steigmann from  UC-Berkeley, Professor C. Horgan from  University of Virginia,
 and Professor Martin Ostoja-Starzewski from University of Illinois are sincerely acknowledged.
%Special thanks to Professor Ray Ogden from Glasgow University for his detailed comments and constructive suggestions.
Reviewer's important comments and constructive suggestions   are sincerely acknowledged.
The research   was supported by US Air Force Office of Scientific Research (AFOSR FA9550-10-1-0487).

\end{document}